\documentclass[%
 reprint,
 amsmath,amssymb,
 aps,
]{revtex4-2}
\usepackage[T1]{fontenc}
\usepackage{graphicx}
\usepackage{dcolumn}
\usepackage{bm}
\usepackage{float}
\usepackage{mathptmx}
\usepackage{eucal}
\usepackage{amssymb}
\usepackage{verbatim}
\usepackage{multirow}
\usepackage{silence}
\usepackage[colorlinks,linkcolor=blue,anchorcolor=blue,citecolor=blue,urlcolor=blue]{hyperref}

\begin{document}

\preprint{APS/123-QED}

\title{Manifold Learning for Dimensionality Reduction: Quantum Isomap algorithm}
\author{WeiJun Feng$,^1$ GongDe Guo$,^1$Kai Yu$,^1$Xin Zhang$,^1$}
\author{Song Lin$^{1,}$}
\thanks{Corresponding author:lins95@gmail.com}%
\affiliation{$^1$College of computer and Cyber Security, Fujian Normal University, Fuzhou 350117, China
}%
\begin{abstract}
Isomap algorithm is a representative manifold learning algorithm. The algorithm simplifies the data analysis process and is widely used in neuroimaging, spectral analysis and other fields. However, the classic Isomap algorithm becomes unwieldy when dealing with large data sets. Our object is to accelerate the classical algorithm with quantum computing, and propose the quantum Isomap algorithm. The algorithm consists of two sub-algorithms. The first one is the quantum Floyd algorithm, which calculates the shortest distance for any two nodes. The other is quantum Isomap algorithm based on quantum Floyd algorithm, which finds a low-dimensional representation for the original high-dimensional data. Finally, we analyze that the quantum Floyd algorithm achieves exponential speedup without sampling. In addition, the time complexity of quantum Isomap algorithm is $O(dNpolylogN)$. Both algorithms reduce the time complexity of classical algorithms.
\end{abstract}
\maketitle
\section{introduction}

\indent With the continuous popularization of "digitalization", a large amount of information produced in real life is digitized. In the digital age, it has become essential to store content such as text, webpages and images with fewer resources. The challenge of using high-dimensional data for machine learning tasks, moreover, are not conducive to training. Among problems are inescapable for all machine learning methods, which is called the "curse of dimensionality" \cite{ref1}. For these problems, there are two common solutions. The first one is to increase the number of samples and, thus, improve the density of training set. However, this is infeasible since the complexity is exponential. Another approach is widely accepted, which is called dimensionality reduction \cite{ref2,ref3,ref4}. By filtering out non-essential features of high-dimensional samples, the proposed method can obtain a low-dimensional representation of the original high-dimensional data. In general, dimensionality reduction methods can be categorized as linear or nonlinear methods. The basic idea of the former is to find low-dimensional projections and extract essential information from the data. The latter assumes the original high-dimensional data distributed on the low-dimensional manifold, and reconstructs the original data on the low-dimensional manifold. In addition, these methods are widely used in fields such as pattern recognition, image processing and text analysis.\\
\indent Undoubtedly, dimensionality reduction algorithms address data storage and improve the learning efficiency of machine learning models. However, this algorithm yields a high time complexity. Given a data matrix of $N\times D$, where $N$ and $D$ represent the number of samples and the dimension of samples, respectively, the time complexity is $O(N^3)$. On the other hand, quantum computation is one of the important areas of quantum information technology. Since the propositions of Shor algorithm \cite{ref5} and Grover algorithm \cite{ref6}, the quantum computing \cite{ref7,ref8} have attracted much attention. Recently, there has been a lot of research on combining machine learning tasks with quantum computing. This further stimulates us to investigate to reduce the time complexity of dimension reduction algorithms with quantum computing.\\
\indent So far, researchers have proposed some quantum versions of dimensionality reduction algorithms. In 2014, Professor Loyd's group \cite{ref9} came up with the first dimensional reduction algorithm, quantum principal component analysis, which provided a reference for many subsequent quantum algorithms. The algorithm aims to maximize the variance of the projected samples using the projection matrix, and eventually achieve dimensionality reduction. Formally, quantum principal component analysis algorithm achieves the exponential acceleration compared with the classical algorithm counterpart. In 2016, Cong et al. \cite{ref10} proposed a quantum linear discriminant algorithm. The algorithm calculates the ratio of between-class scatter degree to within-class scatter degree, and seeks the optimal projection direction of the data. The overall time complexity of the algorithm is polylogarithmic in both the number of samples and the dimension. Both quantum algorithms speed up the processing of linear dimensionality reduction tasks. However, linear dimensionality reduction algorithms are powerless when the original data is distributed on the manifold. Then, He et al. \cite{ref11} proposed a quantum version of local linear embeding algorithm in 2020. The algorithm assumes the data exists linear relationship between the data and the locally adjacent points, which remains unchanged the relation after dimensionality reduction. Soon after, Li \cite{ref12} proposed a quantum nonlinear dimensionality reduction algorithm based on arbitrary kernel functions. The idea of the algorithm is to approximate arbitrary kernels using the idea of Taylor series. It should be noted that the algorithm performance will vary greatly when selecting different kernel functions. In 2021, Sornsaeng \cite{ref13} proposed a quantum diffusion mapping algorithm inspired by random walks. In fact, the algorithm converts the sample similarity matrix to the analysis of the transition matrix, and the higher similarity corresponds to the higher transition probability. \\
\indent In contrast to the above facts, performing these nonlinear dimensionality reduction algorithms on manifold data, the effect between data pairs that are far apart cannot be guaranteed. In this way, we propose a quantum Isomap algorithm. Let the algorithm perform on the original dataset, and the relative relationships between all the new data will unchanged. One difficulty of our algorithm is calculating the geodesic distance. To obtain the manifold distance between all data pairs, we propose a quantum Floyd algorithm. It should be noted that the quantum Floyd algorithm must effectively load the adjacency matrix, for example, using the quantum random access memory. As a part of the Isomap algorithm, the quantum Floyd algorithm enables us to efficiently complete the algorithm. In addition, we use the quantum singular value estimation algorithm to estimate the eigen-solution of the inner product matrix. Both algorithms provide a significant acceleration effect.\\
\indent The remainder of the paper is organized as follows: Section \uppercase\expandafter{\romannumeral2} introduces the classical Isomap algorithm and the quantum singular value estimation. Section \uppercase\expandafter{\romannumeral3} gives the quantum Isomap algorithm and its sub-algorithm. Section \uppercase\expandafter{\romannumeral4} analyzes the time complexity of the proposed algorithm. Finally, we give relevant conclusions in Sec.\uppercase\expandafter{\romannumeral5}.

\section{preliminaries}
\indent In this section, we review the basic idea and the main procedure of the classical Isomap algorithm. Then, we introduce the quantum singular value estimate and give some necessary derivation steps.
\subsection{Review of Classical Isomap algorithm}
\indent The Isomap algorithm is a nonlinear dimensionality reduction method. The algorithm takes geodesic distance as the input of MDS algorithm, and calculate the low-dimensional results of high-dimensional data. In this subsection, the classical Isomap algorithm consists of the following three steps.\\
\indent In the first step, we compute the geodesic distance for all data. By using Floyd algorithm \cite{ref14}, we obtain the geodesic distance $d_{ij}$ between any two nodes.\\
\indent The second step is to transform the geodesic distance matrix into an inner product matrix. The idea of the derivation is as follows. We assume that the data $Z\in \mathbb{R}^{N\times d}$ represents the result of dimensionality reduction. Let $K=ZZ^T$ the inner product matrix, where $k_{ij}=z_i {z_j}^T$, then
\begin{equation}
	d_{ij}=||z_i||^2+||z_j||^2-2z_i{z_j}^T=k_{ii}+k_{jj}-2k_{ij}.
\end{equation}
Similar to the PCA algorithm, the dataset after dimensionality reduction is assumed to be centralized. i.e., $\sum_{i}z_{i}=0$. Equivalently, the sum of the row or column of the matrix is 0, that is, $\sum_{i}{k_{ij}}=\sum_{j}{k_{ij}}=0$. We can obtain the following relationship
\begin{equation}
	\begin{aligned}
		d_{i*}&=\frac{1}{N}\sum_{j}{d_{ij}}=\frac{1}{N}tr(K)+k_{jj},\\
		d_{*j}&=\frac{1}{N}\sum_{i}{d_{ij}}=\frac{1}{N}tr(K)+k_{ii},\\
		d_{**}&=\frac{1}{N^2}\sum_{i,j}{d_{ij}}=\frac{2}{N}tr(K).
	\end{aligned}
\end{equation}
Here, $d_{i*}$ and $d_{*j}$ represent the row mean of the distance matrix and the column mean of the geodesic distance matrix, respectively, and $d_{**}$ is the mean of all elements of the geodesic distance matrix.\\
Now, we can take each entry of the inner product matrix and represent it as
\begin{equation}
	k_{ij}=-\frac{1}{2} (d_{ij}-d_{i*}-d_{*j}+d_{**}).
\end{equation}
\indent The last step is the eigen-decomposition for the inner product matrix $K$. Since the matrix $K$ is symmetric positive semidefinite, it can be expressed as
\begin{equation}
	K=V\Lambda V^T,
\end{equation}
where $\Lambda=diag(\lambda_1,\lambda_2,\cdots,\lambda_N)$ is a diagonal matrix composed of eigenvalues, and $\lambda_1\ge\lambda_2\ge\cdots\ge\lambda_N$, $V$ is the eigenvector matrix arranged by the size of the eigenvalues. Equivalently, when select the $d$ largest eigenvalue, the data matrix can be represented as
\begin{equation}
	Z=V_*\Lambda_{*}^{\frac{1}{2}}\in \mathbb{R}^{N\times d},
\end{equation}
where $\Lambda_{*}=diag{(\lambda_1,\lambda_2,\cdots,\lambda_d,0,\cdots,0)}$, $V_*$ is the corresponding eigenvector.
\subsection{Quantum singular value estimation}
\indent Quantum singular value estimation \cite{ref15,ref16} is used to estimate the singular value of a matrix. Given the data structure of a matrix or the ability to prepare a matrix quantum state, we can complete the algorithm. Indeed, the algorithm is developed from a discrete-time Markov chain \cite{ref17,ref18}. Given a stochastic matrix $K$, we define two mappings:
\begin{equation}
	\begin{aligned}
		U_{\mathcal{M}}:|0\rangle|j\rangle\to|K_j\rangle|j\rangle&=\frac{1}{||K_j||}\sum_{i=1}^{M}K_{ij}|i\rangle|j\rangle,\\
		U_{\mathcal{N}}:|i\rangle|0\rangle\to|i\rangle|K_F\rangle&=\frac{1}{||K||_F}\sum_{j=1}^{M}||K_j|||i\rangle|j\rangle,
	\end{aligned}
\end{equation}
where $|K_j\rangle=\sum_{i=1}^{N}K_{ij}/||K_j|||i\rangle$, $|K_F\rangle=\sum_{j=1}^{N}||A_j||/||K||_F|j\rangle$. To quantize the Markov chain, we construct two matrices of the following form:
\begin{equation}
	\begin{aligned}
		\mathcal{M}&=\sum_{j=1}^{N}|K_j\rangle|j\rangle\langle j|,\\
		\mathcal{N}&=\sum_{i=1}^{N}|i\rangle|K_F\rangle\langle i|.
	\end{aligned}
\end{equation}
We prove that the product between $\mathcal{N^{\dagger}}$ and $\mathcal{M}$ satisfies
\begin{equation}
	\begin{aligned}
		{\mathcal{N}}^\dagger\mathcal{M}=&{(\sum_{i=1}^{N}|i\rangle\langle i|\langle K_F|)}{(\sum_{j=1}^{N}|K_j\rangle|j\rangle\langle j|)}\\
		=&\frac{K}{||K||_F},
	\end{aligned}
\end{equation}
which corresponds to the stochastic matrix. This allows us define the following unitary transformation
\begin{equation}
	\begin{aligned}
		W&=	{(2\mathcal{N}\mathcal{N}^{\dagger}-I_{N^2})}{(2\mathcal{M}\mathcal{M}^{\dagger}-I_{N^2})}.		
	\end{aligned}
\end{equation}
In order to obtain the spectrum of $K$, we first isolate an invariant subspace for $W$ and look for eigenvalues and eigenvectors in this space. Let us see the effect of $W$ on $\mathcal{M}|v_k\rangle$
\begin{equation}
	\begin{aligned}
		W\mathcal{M}|v_k\rangle&={(2\mathcal{N}\mathcal{N}^{\dagger}-I_{N^2})}{(2\mathcal{M}\mathcal{M}^{\dagger}-I_{N^2})}\mathcal{M}|v_k\rangle\\
		&=\frac{2\sigma_k}{||K||_F}\mathcal{N}|u_k\rangle-\mathcal{M}|v_k\rangle,
	\end{aligned}
\end{equation}
and
\begin{equation}
	W\mathcal{N}|u_k\rangle=(\frac{4\sigma_{k}^{2}}{||K||_{F}^{2}}-1)\mathcal{N}|u_k\rangle-\frac{2\sigma_{k}}{||K||_F}\mathcal{M}|v_k\rangle,
\end{equation}
where $|u_k\rangle$ and $|v_k\rangle$ are the left eigenvector and right eigenvector of $K$, respectively. From (10) and (11), we can deduce that the subspace
\begin{equation}
	\mathcal{I}_W:=\{\mathcal{M}|v_k\rangle,\mathcal{N}|u_k\rangle\}
\end{equation}
is invariant under $W$. Then we orthogonalize the basis using the Gram-Schmidt method \cite{ref19}, and get the result
\begin{equation}
	\begin{aligned}
		|e_{k1}\rangle&=\mathcal{M}|v_k\rangle,\\
		|e_{k2}\rangle&=\frac{\mathcal{N}|u_k\rangle-\sigma_k||K||_{F}^{-1}|e_{k1}\rangle}{\sqrt{1-\sigma_{k}^{2}||K||_{F}^{-2}}}.
	\end{aligned}
\end{equation}
By performing $W$ on $|e_{k1}\rangle$, it leads to
\begin{equation}
	W|e_{k1}\rangle={[\frac{2\sigma_{k}^{2}}{||K||_{F}^{2}}-1]|e_{k1}\rangle}+\frac{2\sigma_k}{||K||_F}\sqrt{1-\frac{\sigma_{k}^{2}}{||K||_{F}^{2}}}|e_{k2}\rangle.
\end{equation}
We consider that $\mathcal{M}|v_k\rangle$ and $\mathcal{N}|u_k\rangle$ are linearly independent \cite{ref20}. Setting that $W$ rotates the basis $|e_{k1}\rangle$ in a fixed angle $2\theta$, the eigenvalue of $W$ will be $e^{2i\theta_{k}}$, and the corresponding eigenvector is $|x_{\pm}^{k}\rangle=1/\sqrt{2}(|e_{k1}\rangle\pm i|e_{k2}\rangle)$. In addition, we have the following equation relation
\begin{equation}
	\begin{aligned}
		cos2\theta_k&=\langle v_k|\mathcal{M}^{\dagger}W\mathcal{M}|v_k\rangle=2\frac{\sigma_{k}^{2}}{||K||_{F}^{2}}-1.
	\end{aligned}
\end{equation}
From the above expression, we obtain the relation $cos\theta_k=\sigma_k/{||K||_{F}^{2}}$, which allows us to solve the eigen-solution using quantum singular value estimation.
\section{Quantum Isomap algorithm}
\subsection{Quantum Floyd algorithm}
\indent The Floyd method is an algorithm for finding the shortest path between any two points. Since the high time complexity of the Floyd method, we propose a quantum variant of the Floyd method. When processing multi-class manifold task \cite{ref21} with a random sampling scheme, we achieve exponential acceleration compared with the classical Floyd algorithm. The procedure of our algorithm is described in detail as follows steps:\\
\indent Step A1: Let the $O_{d_{ij}}$ load the distance information of the nearest neighbor pairs for dataset. Then, we prepare the initial state
\begin{equation}
	|k\rangle\sum_{i,j=1}^{N}|i\rangle_1|j\rangle_2|d_{ij}\rangle_3|0\rangle_4|0\rangle_5|0\rangle_{anc}|0\rangle_6|0\rangle_7|0\rangle_8|0\rangle_9|0\rangle_{10},
\end{equation}
where $d_{ij}$ is the element of the adjacency matrix, and the infinity value of the adjacency matrix is replaced by the larger value. Here, the value of $k$ represents the transit node of the Floyd algorithm. Moreover, we assume the length of the quantum register storing data is $l$, which is sufficient to represent the largest data.\\
\indent Step A2: Using the quantum equality mapping function \cite{ref22} to judge whether $k$ is equal to $i$ or $j$, respectively. Then, we will get the quantum state
\begin{widetext}
	\begin{equation}
		|k\rangle\sum_{i,j=1}^{N}|i\rangle_1|j\rangle_2|d_{ij}\rangle_3|0\rangle_4|0\rangle_5|0\rangle_{anc}(|0\rangle_6|0\rangle_7+|1\rangle_6|0\rangle_7)(|0\rangle_8|0\rangle_9+|1\rangle_8|0\rangle_9)|0\rangle_{10},
	\end{equation}
	where the value 0 in the quantum register 6,8 represents $i,j$ is equal to $k$.\\
	\indent Step A3: Controling the preparation of data information by using quantum registers 6 and 8, and loading information for quantum register 4. Then, we get the quantum state
	\begin{equation}
		|k\rangle\sum_{i,j=1}^{N}|i\rangle_1|j\rangle_2|d_{ij}\rangle_3|d_{ij}\rangle_4|0\rangle_5|0\rangle_{anc}(|0\rangle_6|d_{ik}\rangle_7+|1\rangle_6|11\cdots1\rangle_7)(|0\rangle_8|d_{kj}\rangle_9+|1\rangle_8|11\cdots1\rangle_9)|0\rangle_{10}.
	\end{equation}
	If it is 0 in quantum registers 6, we perform $CNOT$ gate on quantum register 3 and 7. Otherwise, we perform the gate $X$ on quantum register 7. Similarly, the quantum registers 8 and 9 execute the same operations as quantum registers 6 and 7. The setting of $11\cdots1$ in quantum registers 7 and 9 is a trick. On the one hand, the setting can assist the quantum comparison process; on the other hand, it is easy to implement.\\
	\indent Step A4: Considering the quantum registers anc,6 and 7 as a whole, and the quantum registers 8,9 as a whole. By applying a quantum adder \cite{ref23,ref24}, the quantum registers 6 and 7 will become
	\begin{equation}
		|k\rangle\sum_{i,j=1}^{N}|i\rangle_1|j\rangle_2|d_{ij}\rangle_3|d_{ij}\rangle_4|0\rangle_5(|d_{ik}+d_{kj}\rangle+|G\rangle)_{anc,6,7}(|0\rangle|d_{kj}\rangle+|1\rangle|11\cdots1\rangle)_{8,9}|0\rangle_{10},
	\end{equation}
	where the quantum state $|G\rangle$ represents the set of quantum states whose quantum register anc is 1, which is called garbage state.\\
	\indent Step A5: Copying the value of $d_{ik}+d_{kj}$ to quantum register 10 when the anc qubit is 0. At the same time, perform $X$ gate operation on quantum register 5 to prepare for step A6. Then, We will get the quantum state
	\begin{equation}
		|k\rangle\sum_{i,j=1}^{N}|i\rangle_1|j\rangle_2|d_{ij}\rangle_3|d_{ij}\rangle_4|1\rangle_5(|d_{ik}+d_{kj}\rangle+|G\rangle)_{anc,6,7}(|0\rangle|d_{kj}\rangle+|1\rangle|11\cdots1\rangle)_{8,9}|d_{ik}+d_{kj}\rangle_{10},
	\end{equation}
	\indent Step A6: Considering the quantum register 5,anc,6 and 7 as a whole, we then perform the quantum subtraction \cite{ref25} with the quantum register 4. It will be the following state
	\begin{equation}
		|k\rangle\sum_{i,j=1}^{N}|i\rangle_1|j\rangle_2|d_{ij}\rangle_3|d_{ij}\rangle_4(\sum_{d_{ij}>d_{ik}+d_{kj}}|0\rangle_5|1,(d_{ik}+d_{kj})-0,d_{ij}\rangle_{anc,6,7}+|G_1\rangle+|G_2\rangle)(|0\rangle|d_{kj}\rangle+|1\rangle|11\cdots1\rangle)_{8,9}|d_{ik}+d_{kj}\rangle_{10}.
	\end{equation}
	Here, the quantum states $|G_1\rangle$ and $|G_2\rangle$ are both 1 in quantum register 5, and the quantum register 5 is 0 when $d_{ij}>d_{ik}+d_{kj}$. Actually, the above implements a quantum comparator \cite{ref26,ref27,ref28}, which is mainly used to determine whether distance information needs to be updated. Compared with other quantum comparators, the greatest advantage of this scheme is that it can solve the entanglement with the marked registers 5.\\
	\indent Step A7: Inversing the quantum register 4 when the quantum register 5 is 0. Then, copy the information of quantum register 10 to quantum register 4, and we will get the quantum state
	\begin{equation}
		\begin{aligned}
			|k\rangle\sum_{i,j=1}^{N}|i\rangle_1|j\rangle_2|d_{ij}\rangle_3(\sum_{d_{ij}>d_{ik}+d_{kj}}|d_{ik}+d_{kj}\rangle_4|0\rangle_5|1,(d_{ik}+d_{kj})-0,d_{ij}\rangle_{anc,6,7}+|d_{ij}\rangle_4(|G_1\rangle+|G_2\rangle))(|0\rangle|d_{kj}\rangle+|1\rangle\\|11\cdots1\rangle)_{8,9}|d_{ik}+d_{kj}\rangle_{10}
		\end{aligned}
	\end{equation}	
Note that when $d_{ij}>d_{ik}+d_{kj}$, the quantum register anc needs to be 1 to prevent overflow.\\
	\begin{figure}[H]
		\includegraphics[width=0.95\textwidth]{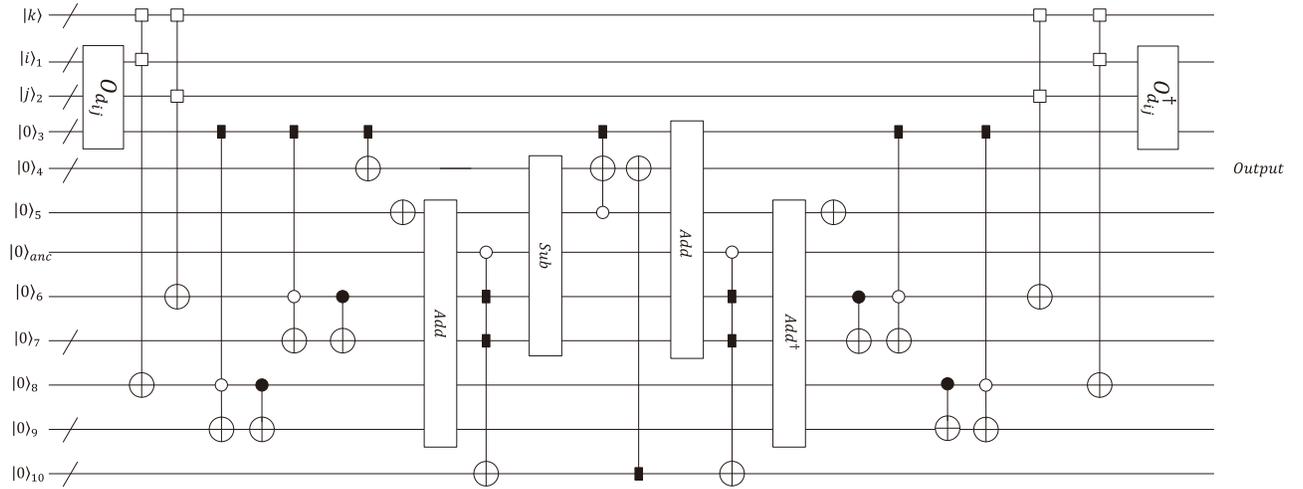}
		\centering
		\caption{\label{1}The first iteration process of quantum Floyd algorithm circuit diagram. The white box represents the map function to check equality, and the long black box indicates the preparation of a copy of the quantum register.}
	\end{figure}
\end{widetext}
\indent Step A8: Performing some inverse operations on Eq. (22), we get the result of the first update
\begin{equation}
	|k\rangle\sum_{i,j=1}^{N}|i\rangle|j\rangle|d_{ij}^{(1)}\rangle,
\end{equation}
where $|d_{ij}^{(1)}\rangle=\sum_{d_{ij}\le d_{ik}+d_{kj}}|d_{ij}\rangle+\sum_{d_{ij}>d_{ik}+d_{kj}}|d_{ik}+d_{kj}\rangle$ represents the distance superposition quantum state. In short, we have achieved the following transformation:
\begin{equation}
	\sum_{i,j=1}^{N}|i\rangle|j\rangle|d_{ij}\rangle\to\sum_{i,j=1}^{N}|i\rangle|j\rangle|d_{ij}^{(1)}\rangle.
\end{equation}
 We also need to be concerned that the result of the current should be the input of the next update. To make the paper more compact, in the next step we start from the initial input state at the r-th iteration.\\
\indent Step A9: Initializing the input quantum state of r-th iteration process as
\begin{equation}
	|k\rangle\sum_{i,j=1}^{N}|i\rangle|j\rangle|d_{ij}^{(r-1)}\rangle,
\end{equation}
where $k$ represents the label of the transit node in the process, and the label of each update is different.\\
\indent Step A10: Performing a procedure similar to the first iteration, we then obtain the geodesic distance matrix quantum state
\begin{equation}
	\sum_{i,j=1}^{N}|i\rangle|j\rangle|d_{ij}^{(r)}\rangle,
\end{equation}
where $\sum_{i,j=1}^{N}|i\rangle|j\rangle|d_{ij}^{(r)}\rangle$ is the result of quantum Floyd algorithm. Since the subsequent algorithm needs to sample, we need an extra step to transfer the information to the amplitude.\\
\indent Step A11: Using the QDAC algorithm \cite{ref29}, we can get the quantum state
\begin{equation}
	\frac{1}{\sqrt{G}}\sum_{i,j=1}^{N}\sqrt{d_{ij}^{(r)}}|i\rangle|j\rangle|d_{ij}^{(r)}\rangle,
\end{equation}
where $G=\sum_{i,j=1}^{N}{d_{ij}^{(r)}}$. Denote the time complexity of steps $A1-A10$ as $T_g$, the time complexity above will be $T_g/P_1$, where $P_1$ represents the sum of the square of the data mean and the variance. It should be noted that applying the amplitude amplification technique \cite{ref30} reduces the original time complexity of $A1-A11$ from $O(T_g/P_1)$ to $O(T_g/\sqrt{P_1})$. In addition, the time complexity of this algorithm scale to $O(polylogN)$ when random sampling \cite{ref31} for multi-class dataset. 
\subsection{Quantum Isomap algorithm based on quantum Floyd algorithm}
\indent In this section, we mainly describe the process of quantum Isomap algorithm, which is based on the quantum Floyd algorithm. The algorithm will be described in three parts as follows.  
\subsubsection{Constructing the geodesic distance quantum state}
\indent In Subsection A of \uppercase\expandafter{\romannumeral3}, we propose a quantum Floyd algorithm. In addition, the introduced quantum Floyd algorithm allows us to prepare two forms of geodesic distance quantum states. One stores the information in the quantum register, we denote it as 
\begin{equation}
	|D_1\rangle=\frac{1}{N}\sum_{i,j=1}^{N}|i\rangle|j\rangle|d_{ij}^{(r)}\rangle,
\end{equation}
which is used as an input state for the preparation of inner product matrix quantum state. The other is stored in the amplitude of the computational basis. We donote it as
\begin{equation}
	|D_2\rangle=\frac{1}{\sqrt{G}}\sum_{i,j=1}^{N}\sqrt{d_{ij}^{(r)}}|i\rangle_1|j\rangle_2{(|d_{ij}^{(r)}\rangle)}_3,
\end{equation}
where $G=\sum_{i,j=1}^{N}{d_{ij}^{(r)}}$. Note that the quantum state $|D_2\rangle$ is used for sampling. Both forms of quantum states allow us to prepare inner product matrix quantum state.\\
\subsubsection{Constructing the inner product matrix quantum state}
\indent According to Eq. (3), we need to calculate the expected values of arbitrary label before preparing the inner product matrix quantum state. We have a quantum model $d_{i*}$, which computes the expectation for any label $i$, namely
\begin{equation}
	d_{i*}=\sum_{j=1}^{N}\langle 0| U^{\dagger}{(\theta)}M_{ij} U{(\theta)}|0\rangle*(d_{ij}^{(r)}),
\end{equation}
where $U{(\theta)}:|0\rangle\mapsto|D_2\rangle$, $M_{ij}:=|i\rangle|j\rangle\langle i|\langle j|\otimes I$ and $G$ represents a constant as shown before. Here, $|d_{i*}-d_{i*}^{(r)}|\le\epsilon$, where $d_{i*}^{(r)}$ is the mean value corresponding to the label $i$, and $\epsilon$ is the error. The time complexity of getting each label' expectation is about $O(T_gNlogN/\sqrt{P_1})$. This enables us to understand each entry by sampling, and then store it in QRAM structure.\\
\indent Now, we begin to describe the process for preparing the inner product matrix quantum state. The steps are as follows:
\indent Step B1: appending quantum registers to the state $|D_1\rangle$, we will get the quantum state
\begin{equation}
	|\phi\rangle_1=\frac{1}{N}\sum_{i,j=1}^{N}|i\rangle_1 |j\rangle_2|d_{ij}^{(r)}\rangle_3|0\rangle_4|0\rangle_5|0\rangle_6|0\rangle_7,
\end{equation}
\indent Step B2: Loading the expected value to the quantum register, and we will get
\begin{equation}
	|\phi\rangle_2=\frac{1}{N}\sum_{i,j=1}^{N}|i\rangle_1 |j\rangle_2{(|d_{ij}^{(r)}\rangle)}_3|d_{i*}\rangle_4|d_{*j}\rangle_5|d_{**}\rangle_6|0\rangle_7,
\end{equation}
Here, $d_{i*}$ and $d_{*j}$ represent the row/column mean of the geodesic distance matrix, and $d_{**}$ can be calculated by $d_{i*}$ and $d_{*j}$ without sampling. Note that the quantum states $|d_{i*}\rangle$ and $|d_{*j}\rangle$ are expressed by 2's complement.\\
\indent Step B3: Executing the quantum adder for quantum registers  3,4,5 and 6, we will the quantum state
\begin{widetext}
	\begin{equation}
			|\phi\rangle_3=\frac{1}{N}\sum_{i,j=1}^{N}|i\rangle_1 |j\rangle_2{|d_{ij}^{(r)}+d_{i*}+d_{*j}+d_{**}\rangle}_3|d_{i*}\rangle_4|d_{*j}\rangle_5|d_{**}\rangle_6|0\rangle_7,
	\end{equation}
where $d_{i*}$ and $d_{*j}$ are 2's complement of $-d_{i*}$ and $-d_{*j}$, respectively. \\
\indent Step B4: Performing the shift-R operation \cite{ref32} on quantum register 3, which is equivalent to multiplying 1/2. Next, we apply one $X$ gate to the quantum register 7 for the next step, the quantum state will be
\begin{equation}
	|\phi\rangle_4=\frac{1}{N}\sum_{i,j=1}^{N}|i\rangle_1 |j\rangle_2{|\frac{1}{2}(d_{ij}^{(r)}+d_{i*}+d_{*j}+d_{**})\rangle}_3|d_{i*}\rangle_4|d_{*j}\rangle_5|d_{**}\rangle_6|1\rangle_7,
\end{equation}
\indent Step B5: Performing $X$ gates on quantum register 3, and then performing a quantum adder on quantum registers 3 and 7. We finally get the quantum state
\begin{equation}
	|\phi\rangle_5=\frac{1}{N}\sum_{i,j=1}^{N}|i\rangle_1 |j\rangle_2{|-\frac{1}{2}(d_{ij}^{(r)}+d_{i*}+d_{*j}+d_{**})\rangle}_3|d_{i*}\rangle_4|d_{*j}\rangle_5|d_{**}\rangle_6|1\rangle_7,
\end{equation} 
\end{widetext}
where the quantum register 3 storing the inner product matrix information.\\
\indent Step B6: Performing some inverse operations on Eq. (35), we get the result
\begin{equation}
	|\phi\rangle_6=\frac{1}{N}\sum_ {i,j=1}^{N}|i\rangle|j\rangle|k{(i,j)}\rangle,
\end{equation}
where $k(i,j)=-\frac{1}{2}(d_{ij}^{(r)}+d_{i*}+d_{*j}+d_{**})$ rerpesents the entry of the inner product matrix.\\
\indent Step B7: Executing the QDAC algorithm, which allows us prepare the state
\begin{equation}
	\begin{aligned}
		|K\rangle=\frac{1}{\sqrt{C}}\sum_{i,j=1}^{N}{k(i,j)}|i\rangle|j\rangle.
	\end{aligned}
\end{equation}
where $C=\sum_{i,j=1}^{N}{k(i,j)}^2$. Let $P_2$ the sum of square of the data mean and variance. Combining with quantum amplifying technology, we can infer that the time complexity of preparing inner product quantum states is $O({T_g/\sqrt{P_2}})$.\\
\begin{figure}[H]
	\includegraphics[width=0.475\textwidth]{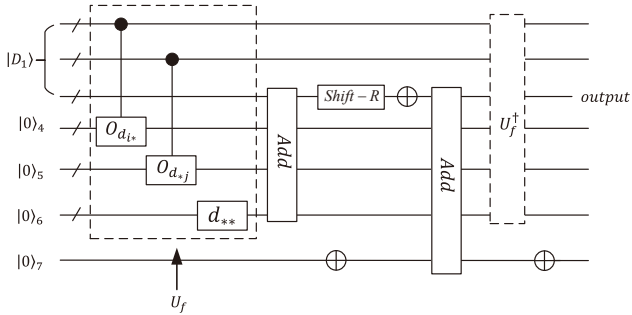}
	\centering
	\caption{\label{3}Preparation of inner product matrix quantum states. The black dot in $U_f$ controls the loading of data. In addition, Shift-R implements: $|a_ma_{m-1}\dots a_2a_10\rangle \to|0a_ma_{m-1}\dots a_1\rangle $.}
\end{figure}
\subsubsection{Eigensolution of the inner product matrix}
\indent The last step of the quantum isomap algorithm is to solve the eigensolution for the inner product matrix. As previously mentioned, the inner product matrix $K$ is a symmetric positive semidefinite. Let the inner product matrix $K=\sum_{k=1}^{N}\sigma_k|v_k\rangle\langle v_k|$, we rewrite the inner product quantum state as
\begin{equation}
	|K\rangle=\sum_{k=1}^{N}\sigma_k|v_k\rangle|v_k\rangle.
\end{equation}
In quantum singular value estimation of Section \uppercase\expandafter{\romannumeral2}, we mentioned that the ability to prepare a matrix quantum state allows us efficient prepare the unitary $U_{\mathcal{M}}$ and $U_{\mathcal{N}}$. Then, we perform quantum singular value estimation on $|K\rangle$ as the following steps:\\
\indent Step C1: Appending $\lceil logN \rceil$ qubits on the state $K$, we perform the map $\mathcal{M}$ on the second register of $|K\rangle$ to obtain the state
\begin{equation}
	\begin{aligned}
		|\psi\rangle_1&=\sum_{k=1}^{N}\sigma_k|v_k\rangle\mathcal{M}|v_k\rangle
		\\&=\sum_{k=1}^{N}\sigma_k|v_k\rangle \sqrt{2}{(|x_{+}^{k}\rangle+|x_{-}^{k}\rangle)}		
	\end{aligned}
	\end{equation}
where $|x_{\pm}^{k}\rangle$ represents the eigenvector of the operation $W$.\\
\indent Step C2: Further append some qubits, and performing the quantum phase estimation algorithm, and obtain the result
  \begin{equation}
  	|\psi\rangle_2=\sum_{k=1}^{N}\sigma_k|v_k\rangle\sqrt{2}{(|x_{+}^{k}\rangle|\overline{\theta_k}\rangle+|x_{-}^{k}\rangle|-\overline{\theta_k}\rangle)}.
  \end{equation}
The size of the last quantum register depends on the accuracy $\epsilon$ of the circuit, which affects the time complexity of the quantum phase estimation algorithm.\\
\indent Step C3: Using an oracle $U_c:|\pm\overline{\theta_k}\rangle|0\rangle\mapsto|\pm\overline{\theta_{k}}\rangle|cos\overline{\theta_{k}}\rangle$, which computes the eigenvalue of the inner product matrix. Storing the $cos\theta_{k}$ in the new quantum register, it leads to the state
  \begin{equation}
	|\psi\rangle_3=\sum_{k=1}^{N}\sigma_k|v_k\rangle\sqrt{2}{(|x_{+}^{k}\rangle|\overline{\theta_k}\rangle+|x_{-}^{k}\rangle|-\overline{\theta_k}\rangle)}|cos\overline{\theta_k}\rangle.
\end{equation}
Note that Wang \cite{ref32} gives one concrete quantum circuit to achieve the oracle $U_c$.\\
\indent Step C4: Undo the quantum phase estimation algorithm, we then obtain
  \begin{equation}
	|\psi\rangle_4=\sum_{k=1}^{N}\sigma_k|v_k\rangle\mathcal{N}|v_k\rangle|cos\overline{\theta_k}\rangle.
\end{equation}
\indent Step C5: Applying the operation $\mathcal{N}^\dagger$, we obtain the state 
  \begin{equation}
	|\psi\rangle_5=\sum_{k=1}^{N}\sigma_k|v_k\rangle|v_k\rangle|cos\overline{\theta_k}\rangle,
\end{equation}
where $cos\overline{\theta_k}$ is approximates to the eigenvalue $\sigma_k$ with accuracy $\epsilon$.
\begin{figure}[H]
	\includegraphics[width=0.475\textwidth]{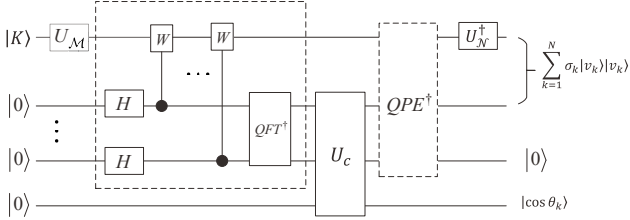}%
	\centering
	\caption{\label{4}A simple quantum circuit describing the eigen-decomposition problem. The dashed box on the left is the unrolled representation of the quantum phase estimation circuit, and $QPE^{\dagger}$ is the inverse phase estimation. $U_c$ represents the quantum circuit for preparing the quantum cosine function.}
\end{figure}
\indent Finally, we sample the state of Eq. (43) by measuring the last register to reveal
the singular values $\sigma_k$, and the corresponding states $|v_k\rangle$. Apparently, we aim to select $d$ largest eigenvalues for the inner product matrix, and $\sigma_k$ satisfy $\sum_{k=1}^{d}\sigma_{k}^{2}/\sum_{k=1}^{N}\sigma_{k}^{2}\ge\eta\approx1$, where $\eta$ is the threshold. Therefore, sampling the state $|\psi_5\rangle$ for $O(d)$ times suffices to obtain $\sigma_k$ and $|v_k\rangle$.\\
\indent To understand the structure of the eigenvector, we can use vector state tomography algorithm \cite{ref33}. Since each $|v_k\rangle$ constituted by the computational basis, this means
\begin{equation}
	|v_k\rangle=\alpha_0|0\rangle+\alpha_1|1\rangle+\cdots+\alpha_{logN-1}|logN-1\rangle.
\end{equation}  
Measure $O(NlogN)$ copies of $|v_k\rangle$ in the standard basis and obtain estimates $\alpha_i=n_i/N$, where $n_i$ is the number of times outcomes $i$ is observed. As a result, we will get the $d$ largest eigenvalues and their corresponding eigenvectors. 
\section{Runtime analysis}
\indent Let us start discussing the time complexity of the whole algorithm. An overview of the time complexity of each step is listed in Table \uppercase\expandafter{\romannumeral1}. More detailed analysis of the algorithm is depicted as follows.\\
\indent For multi-class data sets, the complexity of the quantum Floyd algorithm can be reduced to multilogarithm by using random sampling technology. In fact, we need to sample the quantum circuit before the Steps $B1-B7$. Let the complexity of steps $A1-A11$ be $O(T_g/\sqrt{P_1})$, which implies that the sample time complexity is $O(N^2logNT_g/\sqrt{P_1})$. Let the time complexity of $T_g$ be $O(polylogN)$, and $\sqrt{P_1}$ is independent of the dataset size $N$, the time complexity reduced to $O(N^2polylogN)$.\\
\indent Storing each entry in QRAM data structure after sampling, which allows us efficiently prepare the inner product matrix quantum state $|D_1\rangle$ in subsequent step. In Fig. 2, we find that the time complexity of each step is at most $O(polylogN)$. In this way, the complexity of Steps $B1-B7$ are $O((polylogN)/\sqrt{p_2})$. In addition, the data structure of QRAM enables us to construct the unitary $U_{\mathcal{M}}$ and $U_{\mathcal{N}}$ with time complexity $O(polylogN)$. It should be noted that the time complexity of operator $W$ is mainly contributed by $U_{\mathcal{M}}$ and $U_{\mathcal{M}}$. This means the time complexity of steps $C1-C5$ are $O(polylogN/\epsilon\sqrt{P_2})$. We use the state tomography algorithm to extract the eigenvectors for $d$ largest eigenvalues. The time complexity of state tomography algorithm is $O(dNlogN)$.\\
\indent In summary, the quantum Floyd algorithm can exponentially speed up classical Floyd algorithms. But, it will be $O(N^2polylogN)$ after sampling. In this condition, our algorithm provides substantial speed compared to the classical Floyd algorithm that runs in time $O(N^3)$, particularly for the case of large datasets. Furthermore, putting all runtime together, our algorithm has the overall runtime $O(dNpolylogN/\epsilon\sqrt{P_2})$. When the $\epsilon$ and $\sqrt{P_2}$ are independent of the size $N$, the expected runtime reduces to $O(dNpolylogN)$. Compared with the classical time complexity $O(N^3)$ according to Ref. \cite{ref4}, our algorithm produces a significant effect.
\begin{table}[H]
	\caption{\label{tab:table2}Time complexity each step of the quantum Isomap algorithm.}
	\begin{ruledtabular}
		\begin{tabular}{ll}
			\textbf{Steps} & \textbf{Time complexity} \\
			\hline
			$A1-A11$            &$O(T_g/\sqrt{P_1})$\\
			$B1-B7$            &$O((T_g/\sqrt{P_1}+polylogN)/\sqrt{p_2})$\\
			$C1-C5$            &$O(polylogN/\epsilon\sqrt{P_2})$\\
			 State tomography algorithm          &$O(dNpolylogN)$ 			
		\end{tabular}
	\end{ruledtabular}
\end{table}
\section{conclusions}
\indent The Isomap algorithm has many practical applications in real life. As an improvement over the classical Isomap algorithm, quantum Isomap has the significant advantage of lower time complexity. It is particularly involved in the processing of big data samples, which leads to more cost savings. The quantum Floyd algorithm mentioned in is also a highlight of our algorithm, which solves the shortest path for any two data. In our algorithm, the biggest problem in solving the eigensolution of inner product matrix is that the density matrix cannot be prepared by using the partial trace. Then, we adopt the quantum singular value estimation algorithm, which computes the eigenvalue of the inner product matrix in the new invariant subspace. Note that the algorithm relies on the binary structure of the QRAM model. Then, we can efficiently construct the operation for quantum singular value estimation. After analysis, we draw the conclusion that both two algorithms in this paper realize the acceleration of the classical algorithm. Finally, we hope that our algorithm will promote the development of quantum machine learning algorithms and enlighten more readers.
\nocite{*}
\bibliography{apssamp}

\begin{thebibliography}{33}%
\makeatletter
\providecommand \@ifxundefined [1]{%
 \@ifx{#1\undefined}
}%
\providecommand \@ifnum [1]{%
 \ifnum #1\expandafter \@firstoftwo
 \else \expandafter \@secondoftwo
 \fi
}%
\providecommand \@ifx [1]{%
 \ifx #1\expandafter \@firstoftwo
 \else \expandafter \@secondoftwo
 \fi
}%
\providecommand \natexlab [1]{#1}%
\providecommand \enquote  [1]{``#1''}%
\providecommand \bibnamefont  [1]{#1}%
\providecommand \bibfnamefont [1]{#1}%
\providecommand \citenamefont [1]{#1}%
\providecommand \href@noop [0]{\@secondoftwo}%
\providecommand \href [0]{\begingroup \@sanitize@url \@href}%
\providecommand \@href[1]{\@@startlink{#1}\@@href}%
\providecommand \@@href[1]{\endgroup#1\@@endlink}%
\providecommand \@sanitize@url [0]{\catcode `\\12\catcode `\$12\catcode
  `\&12\catcode `\#12\catcode `\^12\catcode `\_12\catcode `\%12\relax}%
\providecommand \@@startlink[1]{}%
\providecommand \@@endlink[0]{}%
\providecommand \url  [0]{\begingroup\@sanitize@url \@url }%
\providecommand \@url [1]{\endgroup\@href {#1}{\urlprefix }}%
\providecommand \urlprefix  [0]{URL }%
\providecommand \Eprint [0]{\href }%
\providecommand \doibase [0]{https://doi.org/}%
\providecommand \selectlanguage [0]{\@gobble}%
\providecommand \bibinfo  [0]{\@secondoftwo}%
\providecommand \bibfield  [0]{\@secondoftwo}%
\providecommand \translation [1]{[#1]}%
\providecommand \BibitemOpen [0]{}%
\providecommand \bibitemStop [0]{}%
\providecommand \bibitemNoStop [0]{.\EOS\space}%
\providecommand \EOS [0]{\spacefactor3000\relax}%
\providecommand \BibitemShut  [1]{\csname bibitem#1\endcsname}%
\let\auto@bib@innerbib\@empty
\bibitem [{\citenamefont {Indyk}\ and\ \citenamefont {Motwani}(1998)}]{ref1}%
  \BibitemOpen
  \bibfield  {author} {\bibinfo {author} {\bibfnamefont {P.}~\bibnamefont
  {Indyk}}\ and\ \bibinfo {author} {\bibfnamefont {R.}~\bibnamefont
  {Motwani}},\ }\bibfield  {title} {\bibinfo {title} {Approximate nearest
  neighbors: towards removing the curse of dimensionality},\ }in\ \href@noop {}
  {\emph {\bibinfo {booktitle} {Proceedings of the thirtieth annual ACM
  symposium on Theory of computing}}}\ (\bibinfo {year} {1998})\ pp.\ \bibinfo
  {pages} {604--613}\BibitemShut {NoStop}%
\bibitem [{\citenamefont {Ma}\ and\ \citenamefont {Zhu}(2013)}]{ref2}%
  \BibitemOpen
  \bibfield  {author} {\bibinfo {author} {\bibfnamefont {Y.}~\bibnamefont
  {Ma}}\ and\ \bibinfo {author} {\bibfnamefont {L.}~\bibnamefont {Zhu}},\
  }\bibfield  {title} {\bibinfo {title} {A review on dimension reduction},\
  }\href@noop {} {\bibfield  {journal} {\bibinfo  {journal} {International
  Statistical Review}\ }\textbf {\bibinfo {volume} {81}},\ \bibinfo {pages}
  {134} (\bibinfo {year} {2013})}\BibitemShut {NoStop}%
\bibitem [{\citenamefont {Espadoto}\ \emph {et~al.}(2019)\citenamefont
  {Espadoto}, \citenamefont {Martins}, \citenamefont {Kerren}, \citenamefont
  {Hirata},\ and\ \citenamefont {Telea}}]{ref3}%
  \BibitemOpen
  \bibfield  {author} {\bibinfo {author} {\bibfnamefont {M.}~\bibnamefont
  {Espadoto}}, \bibinfo {author} {\bibfnamefont {R.~M.}\ \bibnamefont
  {Martins}}, \bibinfo {author} {\bibfnamefont {A.}~\bibnamefont {Kerren}},
  \bibinfo {author} {\bibfnamefont {N.~S.}\ \bibnamefont {Hirata}},\ and\
  \bibinfo {author} {\bibfnamefont {A.~C.}\ \bibnamefont {Telea}},\ }\bibfield
  {title} {\bibinfo {title} {Toward a quantitative survey of dimension
  reduction techniques},\ }\href@noop {} {\bibfield  {journal} {\bibinfo
  {journal} {IEEE transactions on visualization and computer graphics}\
  }\textbf {\bibinfo {volume} {27}},\ \bibinfo {pages} {2153} (\bibinfo {year}
  {2019})}\BibitemShut {NoStop}%
\bibitem [{\citenamefont {Anowar}\ \emph {et~al.}(2021)\citenamefont {Anowar},
  \citenamefont {Sadaoui},\ and\ \citenamefont {Selim}}]{ref4}%
  \BibitemOpen
  \bibfield  {author} {\bibinfo {author} {\bibfnamefont {F.}~\bibnamefont
  {Anowar}}, \bibinfo {author} {\bibfnamefont {S.}~\bibnamefont {Sadaoui}},\
  and\ \bibinfo {author} {\bibfnamefont {B.}~\bibnamefont {Selim}},\ }\bibfield
   {title} {\bibinfo {title} {Conceptual and empirical comparison of
  dimensionality reduction algorithms (pca, kpca, lda, mds, svd, lle, isomap,
  le, ica, t-sne)},\ }\href@noop {} {\bibfield  {journal} {\bibinfo  {journal}
  {Computer Science Review}\ }\textbf {\bibinfo {volume} {40}},\ \bibinfo
  {pages} {100378} (\bibinfo {year} {2021})}\BibitemShut {NoStop}%
\bibitem [{\citenamefont {LaPierre}(2021)}]{ref5}%
  \BibitemOpen
  \bibfield  {author} {\bibinfo {author} {\bibfnamefont {R.}~\bibnamefont
  {LaPierre}},\ }\bibfield  {title} {\bibinfo {title} {Shor algorithm},\ }in\
  \href@noop {} {\emph {\bibinfo {booktitle} {Introduction to Quantum
  Computing}}}\ (\bibinfo  {publisher} {Springer},\ \bibinfo {year} {2021})\
  pp.\ \bibinfo {pages} {177--192}\BibitemShut {NoStop}%
\bibitem [{\citenamefont {Khanal}\ \emph {et~al.}(2021)\citenamefont {Khanal},
  \citenamefont {Rivas}, \citenamefont {Orduz},\ and\ \citenamefont
  {Zhakubayev}}]{ref6}%
  \BibitemOpen
  \bibfield  {author} {\bibinfo {author} {\bibfnamefont {B.}~\bibnamefont
  {Khanal}}, \bibinfo {author} {\bibfnamefont {P.}~\bibnamefont {Rivas}},
  \bibinfo {author} {\bibfnamefont {J.}~\bibnamefont {Orduz}},\ and\ \bibinfo
  {author} {\bibfnamefont {A.}~\bibnamefont {Zhakubayev}},\ }\bibfield  {title}
  {\bibinfo {title} {Quantum machine learning: A case study of grover’s
  algorithm},\ }in\ \href@noop {} {\emph {\bibinfo {booktitle} {2021
  International Conference on Computational Science and Computational
  Intelligence (CSCI)}}}\ (\bibinfo {organization} {IEEE},\ \bibinfo {year}
  {2021})\ pp.\ \bibinfo {pages} {79--84}\BibitemShut {NoStop}%
\bibitem [{\citenamefont {Alchieri}\ \emph {et~al.}(2021)\citenamefont
  {Alchieri}, \citenamefont {Badalotti}, \citenamefont {Bonardi},\ and\
  \citenamefont {Bianco}}]{ref7}%
  \BibitemOpen
  \bibfield  {author} {\bibinfo {author} {\bibfnamefont {L.}~\bibnamefont
  {Alchieri}}, \bibinfo {author} {\bibfnamefont {D.}~\bibnamefont {Badalotti}},
  \bibinfo {author} {\bibfnamefont {P.}~\bibnamefont {Bonardi}},\ and\ \bibinfo
  {author} {\bibfnamefont {S.}~\bibnamefont {Bianco}},\ }\bibfield  {title}
  {\bibinfo {title} {An introduction to quantum machine learning: from quantum
  logic to quantum deep learning},\ }\href@noop {} {\bibfield  {journal}
  {\bibinfo  {journal} {Quantum Machine Intelligence}\ }\textbf {\bibinfo
  {volume} {3}},\ \bibinfo {pages} {1} (\bibinfo {year} {2021})}\BibitemShut
  {NoStop}%
\bibitem [{\citenamefont {Massoli}\ \emph {et~al.}(2022)\citenamefont
  {Massoli}, \citenamefont {Vadicamo}, \citenamefont {Amato},\ and\
  \citenamefont {Falchi}}]{ref8}%
  \BibitemOpen
  \bibfield  {author} {\bibinfo {author} {\bibfnamefont {F.~V.}\ \bibnamefont
  {Massoli}}, \bibinfo {author} {\bibfnamefont {L.}~\bibnamefont {Vadicamo}},
  \bibinfo {author} {\bibfnamefont {G.}~\bibnamefont {Amato}},\ and\ \bibinfo
  {author} {\bibfnamefont {F.}~\bibnamefont {Falchi}},\ }\bibfield  {title}
  {\bibinfo {title} {A leap among quantum computing and quantum neural
  networks: A survey},\ }\href@noop {} {\bibfield  {journal} {\bibinfo
  {journal} {ACM Computing Surveys (CSUR)}\ } (\bibinfo {year}
  {2022})}\BibitemShut {NoStop}%
\bibitem [{\citenamefont {Lloyd}\ \emph {et~al.}(2014)\citenamefont {Lloyd},
  \citenamefont {Mohseni},\ and\ \citenamefont {Rebentrost}}]{ref9}%
  \BibitemOpen
  \bibfield  {author} {\bibinfo {author} {\bibfnamefont {S.}~\bibnamefont
  {Lloyd}}, \bibinfo {author} {\bibfnamefont {M.}~\bibnamefont {Mohseni}},\
  and\ \bibinfo {author} {\bibfnamefont {P.}~\bibnamefont {Rebentrost}},\
  }\bibfield  {title} {\bibinfo {title} {Quantum principal component
  analysis},\ }\href@noop {} {\bibfield  {journal} {\bibinfo  {journal} {Nature
  Physics}\ }\textbf {\bibinfo {volume} {10}},\ \bibinfo {pages} {631}
  (\bibinfo {year} {2014})}\BibitemShut {NoStop}%
\bibitem [{\citenamefont {Cong}\ and\ \citenamefont {Duan}(2016)}]{ref10}%
  \BibitemOpen
  \bibfield  {author} {\bibinfo {author} {\bibfnamefont {I.}~\bibnamefont
  {Cong}}\ and\ \bibinfo {author} {\bibfnamefont {L.}~\bibnamefont {Duan}},\
  }\bibfield  {title} {\bibinfo {title} {Quantum discriminant analysis for
  dimensionality reduction and classification},\ }\href@noop {} {\bibfield
  {journal} {\bibinfo  {journal} {New Journal of Physics}\ }\textbf {\bibinfo
  {volume} {18}},\ \bibinfo {pages} {073011} (\bibinfo {year}
  {2016})}\BibitemShut {NoStop}%
\bibitem [{\citenamefont {He}\ \emph {et~al.}(2020)\citenamefont {He},
  \citenamefont {Sun}, \citenamefont {Lyu},\ and\ \citenamefont
  {Wang}}]{ref11}%
  \BibitemOpen
  \bibfield  {author} {\bibinfo {author} {\bibfnamefont {X.}~\bibnamefont
  {He}}, \bibinfo {author} {\bibfnamefont {L.}~\bibnamefont {Sun}}, \bibinfo
  {author} {\bibfnamefont {C.}~\bibnamefont {Lyu}},\ and\ \bibinfo {author}
  {\bibfnamefont {X.}~\bibnamefont {Wang}},\ }\bibfield  {title} {\bibinfo
  {title} {Quantum locally linear embedding for nonlinear dimensionality
  reduction},\ }\href@noop {} {\bibfield  {journal} {\bibinfo  {journal}
  {Quantum Information Processing}\ }\textbf {\bibinfo {volume} {19}},\
  \bibinfo {pages} {1} (\bibinfo {year} {2020})}\BibitemShut {NoStop}%
\bibitem [{\citenamefont {Li}\ \emph {et~al.}(2020)\citenamefont {Li},
  \citenamefont {Zhou}, \citenamefont {Xu}, \citenamefont {Hu},\ and\
  \citenamefont {Fan}}]{ref12}%
  \BibitemOpen
  \bibfield  {author} {\bibinfo {author} {\bibfnamefont {Y.}~\bibnamefont
  {Li}}, \bibinfo {author} {\bibfnamefont {R.-G.}\ \bibnamefont {Zhou}},
  \bibinfo {author} {\bibfnamefont {R.}~\bibnamefont {Xu}}, \bibinfo {author}
  {\bibfnamefont {W.}~\bibnamefont {Hu}},\ and\ \bibinfo {author}
  {\bibfnamefont {P.}~\bibnamefont {Fan}},\ }\bibfield  {title} {\bibinfo
  {title} {Quantum algorithm for the nonlinear dimensionality reduction with
  arbitrary kernel},\ }\href@noop {} {\bibfield  {journal} {\bibinfo  {journal}
  {Quantum Science and Technology}\ }\textbf {\bibinfo {volume} {6}},\ \bibinfo
  {pages} {014001} (\bibinfo {year} {2020})}\BibitemShut {NoStop}%
\bibitem [{\citenamefont {Sornsaeng}\ \emph {et~al.}(2021)\citenamefont
  {Sornsaeng}, \citenamefont {Dangniam}, \citenamefont {Palittapongarnpim},\
  and\ \citenamefont {Chotibut}}]{ref13}%
  \BibitemOpen
  \bibfield  {author} {\bibinfo {author} {\bibfnamefont {A.}~\bibnamefont
  {Sornsaeng}}, \bibinfo {author} {\bibfnamefont {N.}~\bibnamefont {Dangniam}},
  \bibinfo {author} {\bibfnamefont {P.}~\bibnamefont {Palittapongarnpim}},\
  and\ \bibinfo {author} {\bibfnamefont {T.}~\bibnamefont {Chotibut}},\
  }\bibfield  {title} {\bibinfo {title} {Quantum diffusion map for nonlinear
  dimensionality reduction},\ }\href@noop {} {\bibfield  {journal} {\bibinfo
  {journal} {Physical Review A}\ }\textbf {\bibinfo {volume} {104}},\ \bibinfo
  {pages} {052410} (\bibinfo {year} {2021})}\BibitemShut {NoStop}%
\bibitem [{\citenamefont {Lyu}\ \emph {et~al.}(2021)\citenamefont {Lyu},
  \citenamefont {Chen}, \citenamefont {Cai},\ and\ \citenamefont
  {Piao}}]{ref14}%
  \BibitemOpen
  \bibfield  {author} {\bibinfo {author} {\bibfnamefont {D.}~\bibnamefont
  {Lyu}}, \bibinfo {author} {\bibfnamefont {Z.}~\bibnamefont {Chen}}, \bibinfo
  {author} {\bibfnamefont {Z.}~\bibnamefont {Cai}},\ and\ \bibinfo {author}
  {\bibfnamefont {S.}~\bibnamefont {Piao}},\ }\bibfield  {title} {\bibinfo
  {title} {Robot path planning by leveraging the graph-encoded floyd
  algorithm},\ }\href@noop {} {\bibfield  {journal} {\bibinfo  {journal}
  {Future Generation Computer Systems}\ }\textbf {\bibinfo {volume} {122}},\
  \bibinfo {pages} {204} (\bibinfo {year} {2021})}\BibitemShut {NoStop}%
\bibitem [{\citenamefont {Wossnig}\ \emph {et~al.}(2017)\citenamefont
  {Wossnig}, \citenamefont {Zhao},\ and\ \citenamefont {Prakash}}]{ref15}%
  \BibitemOpen
  \bibfield  {author} {\bibinfo {author} {\bibfnamefont {L.}~\bibnamefont
  {Wossnig}}, \bibinfo {author} {\bibfnamefont {Z.}~\bibnamefont {Zhao}},\ and\
  \bibinfo {author} {\bibfnamefont {A.}~\bibnamefont {Prakash}},\ }\bibfield
  {title} {\bibinfo {title} {Quantum linear system algorithm for dense
  matrices},\ }\href@noop {} {\bibfield  {journal} {\bibinfo  {journal} {arXiv
  preprint arXiv:1704.06174}\ } (\bibinfo {year} {2017})}\BibitemShut {NoStop}%
\bibitem [{\citenamefont {Shao}\ and\ \citenamefont {Xiang}(2018)}]{ref16}%
  \BibitemOpen
  \bibfield  {author} {\bibinfo {author} {\bibfnamefont {C.}~\bibnamefont
  {Shao}}\ and\ \bibinfo {author} {\bibfnamefont {H.}~\bibnamefont {Xiang}},\
  }\bibfield  {title} {\bibinfo {title} {Quantum circulant preconditioner for a
  linear system of equations},\ }\href@noop {} {\bibfield  {journal} {\bibinfo
  {journal} {Physical Review A}\ }\textbf {\bibinfo {volume} {98}},\ \bibinfo
  {pages} {062321} (\bibinfo {year} {2018})}\BibitemShut {NoStop}%
\bibitem [{\citenamefont {Szegedy}(2004)}]{ref17}%
  \BibitemOpen
  \bibfield  {author} {\bibinfo {author} {\bibfnamefont {M.}~\bibnamefont
  {Szegedy}},\ }\bibfield  {title} {\bibinfo {title} {Quantum speed-up of
  markov chain based algorithms},\ }in\ \href@noop {} {\emph {\bibinfo
  {booktitle} {45th Annual IEEE symposium on foundations of computer
  science}}}\ (\bibinfo {organization} {IEEE},\ \bibinfo {year} {2004})\ pp.\
  \bibinfo {pages} {32--41}\BibitemShut {NoStop}%
\bibitem [{\citenamefont {Paparo}\ and\ \citenamefont
  {Martin-Delgado}(2012)}]{ref18}%
  \BibitemOpen
  \bibfield  {author} {\bibinfo {author} {\bibfnamefont {G.~D.}\ \bibnamefont
  {Paparo}}\ and\ \bibinfo {author} {\bibfnamefont {M.}~\bibnamefont
  {Martin-Delgado}},\ }\bibfield  {title} {\bibinfo {title} {Google in a
  quantum network},\ }\href@noop {} {\bibfield  {journal} {\bibinfo  {journal}
  {Scientific reports}\ }\textbf {\bibinfo {volume} {2}},\ \bibinfo {pages} {1}
  (\bibinfo {year} {2012})}\BibitemShut {NoStop}%
\bibitem [{\citenamefont {Nielsen}\ and\ \citenamefont {Chuang}(2002)}]{ref19}%
  \BibitemOpen
  \bibfield  {author} {\bibinfo {author} {\bibfnamefont {M.~A.}\ \bibnamefont
  {Nielsen}}\ and\ \bibinfo {author} {\bibfnamefont {I.}~\bibnamefont
  {Chuang}},\ }\href@noop {} {\bibinfo {title} {Quantum computation and quantum
  information}} (\bibinfo {year} {2002})\BibitemShut {NoStop}%
\bibitem [{\citenamefont {Liu}\ \emph {et~al.}(2018)\citenamefont {Liu},
  \citenamefont {Yuan}, \citenamefont {Duan},\ and\ \citenamefont
  {Li}}]{ref20}%
  \BibitemOpen
  \bibfield  {author} {\bibinfo {author} {\bibfnamefont {Y.}~\bibnamefont
  {Liu}}, \bibinfo {author} {\bibfnamefont {J.}~\bibnamefont {Yuan}}, \bibinfo
  {author} {\bibfnamefont {B.}~\bibnamefont {Duan}},\ and\ \bibinfo {author}
  {\bibfnamefont {D.}~\bibnamefont {Li}},\ }\bibfield  {title} {\bibinfo
  {title} {Quantum walks on regular uniform hypergraphs},\ }\href@noop {}
  {\bibfield  {journal} {\bibinfo  {journal} {Scientific reports}\ }\textbf
  {\bibinfo {volume} {8}},\ \bibinfo {pages} {1} (\bibinfo {year}
  {2018})}\BibitemShut {NoStop}%
\bibitem [{\citenamefont {Wu}\ and\ \citenamefont {Chan}(2004)}]{ref21}%
  \BibitemOpen
  \bibfield  {author} {\bibinfo {author} {\bibfnamefont {Y.}~\bibnamefont
  {Wu}}\ and\ \bibinfo {author} {\bibfnamefont {K.~L.}\ \bibnamefont {Chan}},\
  }\bibfield  {title} {\bibinfo {title} {An extended isomap algorithm for
  learning multi-class manifold},\ }in\ \href@noop {} {\emph {\bibinfo
  {booktitle} {Proceedings of 2004 International Conference on Machine Learning
  and Cybernetics (IEEE Cat. No. 04EX826)}}},\ Vol.~\bibinfo {volume} {6}\
  (\bibinfo {organization} {IEEE},\ \bibinfo {year} {2004})\ pp.\ \bibinfo
  {pages} {3429--3433}\BibitemShut {NoStop}%
\bibitem [{\citenamefont {Kerenidis}\ and\ \citenamefont
  {Landman}(2021)}]{ref22}%
  \BibitemOpen
  \bibfield  {author} {\bibinfo {author} {\bibfnamefont {I.}~\bibnamefont
  {Kerenidis}}\ and\ \bibinfo {author} {\bibfnamefont {J.}~\bibnamefont
  {Landman}},\ }\bibfield  {title} {\bibinfo {title} {Quantum spectral
  clustering},\ }\href@noop {} {\bibfield  {journal} {\bibinfo  {journal}
  {Physical Review A}\ }\textbf {\bibinfo {volume} {103}},\ \bibinfo {pages}
  {042415} (\bibinfo {year} {2021})}\BibitemShut {NoStop}%
\bibitem [{\citenamefont {Takahashi}\ \emph {et~al.}(2009)\citenamefont
  {Takahashi}, \citenamefont {Tani},\ and\ \citenamefont {Kunihiro}}]{ref23}%
  \BibitemOpen
  \bibfield  {author} {\bibinfo {author} {\bibfnamefont {Y.}~\bibnamefont
  {Takahashi}}, \bibinfo {author} {\bibfnamefont {S.}~\bibnamefont {Tani}},\
  and\ \bibinfo {author} {\bibfnamefont {N.}~\bibnamefont {Kunihiro}},\
  }\bibfield  {title} {\bibinfo {title} {Quantum addition circuits and
  unbounded fan-out},\ }\href@noop {} {\bibfield  {journal} {\bibinfo
  {journal} {arXiv preprint arXiv:0910.2530}\ } (\bibinfo {year}
  {2009})}\BibitemShut {NoStop}%
\bibitem [{\citenamefont {Gidney}(2018)}]{ref24}%
  \BibitemOpen
  \bibfield  {author} {\bibinfo {author} {\bibfnamefont {C.}~\bibnamefont
  {Gidney}},\ }\bibfield  {title} {\bibinfo {title} {Halving the cost of
  quantum addition},\ }\href@noop {} {\bibfield  {journal} {\bibinfo  {journal}
  {Quantum}\ }\textbf {\bibinfo {volume} {2}},\ \bibinfo {pages} {74} (\bibinfo
  {year} {2018})}\BibitemShut {NoStop}%
\bibitem [{\citenamefont {Thapliyal}\ \emph {et~al.}(2019)\citenamefont
  {Thapliyal}, \citenamefont {Munoz-Coreas}, \citenamefont {Varun},\ and\
  \citenamefont {Humble}}]{ref25}%
  \BibitemOpen
  \bibfield  {author} {\bibinfo {author} {\bibfnamefont {H.}~\bibnamefont
  {Thapliyal}}, \bibinfo {author} {\bibfnamefont {E.}~\bibnamefont
  {Munoz-Coreas}}, \bibinfo {author} {\bibfnamefont {T.}~\bibnamefont
  {Varun}},\ and\ \bibinfo {author} {\bibfnamefont {T.~S.}\ \bibnamefont
  {Humble}},\ }\bibfield  {title} {\bibinfo {title} {Quantum circuit designs of
  integer division optimizing t-count and t-depth},\ }\href@noop {} {\bibfield
  {journal} {\bibinfo  {journal} {IEEE transactions on emerging topics in
  computing}\ }\textbf {\bibinfo {volume} {9}},\ \bibinfo {pages} {1045}
  (\bibinfo {year} {2019})}\BibitemShut {NoStop}%
\bibitem [{\citenamefont {Wiebe}\ \emph {et~al.}(2014)\citenamefont {Wiebe},
  \citenamefont {Kapoor},\ and\ \citenamefont {Svore}}]{ref26}%
  \BibitemOpen
  \bibfield  {author} {\bibinfo {author} {\bibfnamefont {N.}~\bibnamefont
  {Wiebe}}, \bibinfo {author} {\bibfnamefont {A.}~\bibnamefont {Kapoor}},\ and\
  \bibinfo {author} {\bibfnamefont {K.}~\bibnamefont {Svore}},\ }\bibfield
  {title} {\bibinfo {title} {Quantum algorithms for nearest-neighbor methods
  for supervised and unsupervised learning},\ }\href@noop {} {\bibfield
  {journal} {\bibinfo  {journal} {arXiv preprint arXiv:1401.2142}\ } (\bibinfo
  {year} {2014})}\BibitemShut {NoStop}%
\bibitem [{\citenamefont {Heidari}\ and\ \citenamefont
  {Farzadnia}(2017)}]{ref27}%
  \BibitemOpen
  \bibfield  {author} {\bibinfo {author} {\bibfnamefont {S.}~\bibnamefont
  {Heidari}}\ and\ \bibinfo {author} {\bibfnamefont {E.}~\bibnamefont
  {Farzadnia}},\ }\bibfield  {title} {\bibinfo {title} {A novel quantum
  lsb-based steganography method using the gray code for colored quantum
  images},\ }\href@noop {} {\bibfield  {journal} {\bibinfo  {journal} {Quantum
  Information Processing}\ }\textbf {\bibinfo {volume} {16}},\ \bibinfo {pages}
  {1} (\bibinfo {year} {2017})}\BibitemShut {NoStop}%
\bibitem [{\citenamefont {Li}\ \emph {et~al.}(2022{\natexlab{a}})\citenamefont
  {Li}, \citenamefont {Lin}, \citenamefont {Yu},\ and\ \citenamefont
  {Guo}}]{ref28}%
  \BibitemOpen
  \bibfield  {author} {\bibinfo {author} {\bibfnamefont {J.}~\bibnamefont
  {Li}}, \bibinfo {author} {\bibfnamefont {S.}~\bibnamefont {Lin}}, \bibinfo
  {author} {\bibfnamefont {K.}~\bibnamefont {Yu}},\ and\ \bibinfo {author}
  {\bibfnamefont {G.}~\bibnamefont {Guo}},\ }\bibfield  {title} {\bibinfo
  {title} {Quantum k-nearest neighbor classification algorithm based on hamming
  distance},\ }\href@noop {} {\bibfield  {journal} {\bibinfo  {journal}
  {Quantum Information Processing}\ }\textbf {\bibinfo {volume} {21}},\
  \bibinfo {pages} {1} (\bibinfo {year} {2022}{\natexlab{a}})}\BibitemShut
  {NoStop}%
\bibitem [{\citenamefont {Mitarai}\ \emph {et~al.}(2019)\citenamefont
  {Mitarai}, \citenamefont {Kitagawa},\ and\ \citenamefont {Fujii}}]{ref29}%
  \BibitemOpen
  \bibfield  {author} {\bibinfo {author} {\bibfnamefont {K.}~\bibnamefont
  {Mitarai}}, \bibinfo {author} {\bibfnamefont {M.}~\bibnamefont {Kitagawa}},\
  and\ \bibinfo {author} {\bibfnamefont {K.}~\bibnamefont {Fujii}},\ }\bibfield
   {title} {\bibinfo {title} {Quantum analog-digital conversion},\ }\href@noop
  {} {\bibfield  {journal} {\bibinfo  {journal} {Physical Review A}\ }\textbf
  {\bibinfo {volume} {99}},\ \bibinfo {pages} {012301} (\bibinfo {year}
  {2019})}\BibitemShut {NoStop}%
\bibitem [{\citenamefont {Brassard}\ \emph {et~al.}(2002)\citenamefont
  {Brassard}, \citenamefont {Hoyer}, \citenamefont {Mosca},\ and\ \citenamefont
  {Tapp}}]{ref30}%
  \BibitemOpen
  \bibfield  {author} {\bibinfo {author} {\bibfnamefont {G.}~\bibnamefont
  {Brassard}}, \bibinfo {author} {\bibfnamefont {P.}~\bibnamefont {Hoyer}},
  \bibinfo {author} {\bibfnamefont {M.}~\bibnamefont {Mosca}},\ and\ \bibinfo
  {author} {\bibfnamefont {A.}~\bibnamefont {Tapp}},\ }\bibfield  {title}
  {\bibinfo {title} {Quantum amplitude amplification and estimation},\
  }\href@noop {} {\bibfield  {journal} {\bibinfo  {journal} {Contemporary
  Mathematics}\ }\textbf {\bibinfo {volume} {305}},\ \bibinfo {pages} {53}
  (\bibinfo {year} {2002})}\BibitemShut {NoStop}%
\bibitem [{\citenamefont {Li}\ \emph {et~al.}(2022{\natexlab{b}})\citenamefont
  {Li}, \citenamefont {Lin}, \citenamefont {Yu},\ and\ \citenamefont
  {Guo}}]{ref31}%
  \BibitemOpen
  \bibfield  {author} {\bibinfo {author} {\bibfnamefont {J.}~\bibnamefont
  {Li}}, \bibinfo {author} {\bibfnamefont {S.}~\bibnamefont {Lin}}, \bibinfo
  {author} {\bibfnamefont {K.}~\bibnamefont {Yu}},\ and\ \bibinfo {author}
  {\bibfnamefont {G.}~\bibnamefont {Guo}},\ }\bibfield  {title} {\bibinfo
  {title} {Quantum k-nearest neighbor classification algorithm based on hamming
  distance},\ }\href@noop {} {\bibfield  {journal} {\bibinfo  {journal}
  {Quantum Information Processing}\ }\textbf {\bibinfo {volume} {21}},\
  \bibinfo {pages} {1} (\bibinfo {year} {2022}{\natexlab{b}})}\BibitemShut
  {NoStop}%
\bibitem [{\citenamefont {Wang}\ \emph {et~al.}(2020)\citenamefont {Wang},
  \citenamefont {Wang}, \citenamefont {Li}, \citenamefont {Fan}, \citenamefont
  {Cui}, \citenamefont {Wei},\ and\ \citenamefont {Gu}}]{ref32}%
  \BibitemOpen
  \bibfield  {author} {\bibinfo {author} {\bibfnamefont {S.}~\bibnamefont
  {Wang}}, \bibinfo {author} {\bibfnamefont {Z.}~\bibnamefont {Wang}}, \bibinfo
  {author} {\bibfnamefont {W.}~\bibnamefont {Li}}, \bibinfo {author}
  {\bibfnamefont {L.}~\bibnamefont {Fan}}, \bibinfo {author} {\bibfnamefont
  {G.}~\bibnamefont {Cui}}, \bibinfo {author} {\bibfnamefont {Z.}~\bibnamefont
  {Wei}},\ and\ \bibinfo {author} {\bibfnamefont {Y.}~\bibnamefont {Gu}},\
  }\bibfield  {title} {\bibinfo {title} {Quantum circuits design for evaluating
  transcendental functions based on a function-value binary expansion method},\
  }\href@noop {} {\bibfield  {journal} {\bibinfo  {journal} {Quantum
  Information Processing}\ }\textbf {\bibinfo {volume} {19}},\ \bibinfo {pages}
  {1} (\bibinfo {year} {2020})}\BibitemShut {NoStop}%
\bibitem [{\citenamefont {Kerenidis}\ and\ \citenamefont
  {Prakash}(2020)}]{ref33}%
  \BibitemOpen
  \bibfield  {author} {\bibinfo {author} {\bibfnamefont {I.}~\bibnamefont
  {Kerenidis}}\ and\ \bibinfo {author} {\bibfnamefont {A.}~\bibnamefont
  {Prakash}},\ }\bibfield  {title} {\bibinfo {title} {A quantum interior point
  method for lps and sdps},\ }\href@noop {} {\bibfield  {journal} {\bibinfo
  {journal} {ACM Transactions on Quantum Computing}\ }\textbf {\bibinfo
  {volume} {1}},\ \bibinfo {pages} {1} (\bibinfo {year} {2020})}\BibitemShut
  {NoStop}%
\end{thebibliography}%
\end{document}